\documentclass[11pt,aps,nofootinbib,prd,aps,epsf,floats,axodraw,amsmath,amssymb,amsfonts]{revtex4}
\usepackage{amsmath, amssymb}
\newcommand{\mathsym}[1]{{}}

\usepackage{graphicx}
\usepackage{amsmath}
\usepackage{amssymb}
\usepackage{bm}
\newcommand{\be}{\begin{equation}}
\newcommand{\ee}{\end{equation}}
\newcommand{\bea}{\begin{eqnarray}}
\newcommand{\eea}{\end{eqnarray}}

\newcommand{\rem}[1]{}
\newsavebox{\PSLASH}
 \sbox{\PSLASH}{$p$\hspace{-1.8mm}/}
 
\renewcommand{\theequation}{\thesection.\arabic{equation}}
\newcounter{saveeqn}
\newcommand{\add}{\addtocounter{equation}{1}}
\newcommand{\alpheqn}{\setcounter{saveeqn}{\value{equation}}%
\setcounter{equation}{0}%
\renewcommand{\theequation}{\mbox{\thesection.\arabic{saveeqn}{\alph{equation}}}}}
\newcommand{\reseteqn}{\setcounter{equation}{\value{saveeqn}}%
\renewcommand{\theequation}{\thesection.\arabic{equation}}}

 \newsavebox{\notrightarrow}
 \sbox{\notrightarrow}{$\to$\hspace{-4mm}/}
 
 \newsavebox{\PARTIALSLASH}
 \sbox{\PARTIALSLASH}{$\partial$\hspace{-1.6mm}/}
 
 \newsavebox{\ASLASH}
 \sbox{\ASLASH}{$A$\hspace{-2.1mm}/}
 
 \newsavebox{\KSLASH}
 \sbox{\KSLASH}{$k$\hspace{-1.8mm}/}
 
 \newsavebox{\LSLASH}
 \sbox{\LSLASH}{$\ell$\hspace{-1.8mm}/}
 
 \newsavebox{\QSLASH}
 \sbox{\QSLASH}{$q$\hspace{-1.8mm}/}
 
 \newsavebox{\DSLASH}
 \sbox{\DSLASH}{$D$\hspace{-2.2mm}/}
 
 \newsavebox{\DbfSLASH}
 \sbox{\DbfSLASH}{${\mathbf D}$\hspace{-2.8mm}/}
 
 \newsavebox{\DELVECRIGHT}
 \sbox{\DELVECRIGHT}{$\stackrel{\rightarrow}{\partial}$}
 
 \newcommand{\blue}{\IfColor{\textCadetBlue}{}}
\newcommand{\black}{\IfColor{\textBlack}{}}
\newcommand{\red}{\IfColor{\textRed}{}}
\newcommand{\green}{\IfColor{\textOliveGreen}{}}
\newcommand{\lila}{\IfColor{\textRedViolet}{}}

\begin{document}
\begin{flushright}
 [math-ph]
\end{flushright}
\title{$\star$-Cohomology, Connes-Chern Characters, and Anomalies in \\General Translation-Invariant Noncommutative Yang-Mills}

\author{Amir Abbass Varshovi}\email{ab.varshovi@sci.ui.ac.ir/amirabbassv@ipm.ir/varshoviamirabbass@gmail.com}

\affiliation{
   Department of Mathematics, University of Isfahan, Isfahan, IRAN.\\
   School of Mathematics, Institute for Research in Fundamental Sciences (IPM), Tehran, IRAN.}
\begin{abstract}
       \textbf{Abstract\textbf{:}} Topological structure of translation-invariant noncommutative Yang-Mills theoris are studied by means of a cohomology theory, so called $\star$-cohomology, which plays an intermediate role between de Rham and cyclic (co)homology theory for noncommutative algebras and gives rise to a cohomological formulation comparable to Seiberg-Witten map. \\
       
\noindent \textbf{Keywords\textbf{:}} Translation-Invariant Star Product, Noncommutative Yang-Mills, Spectral Triple, Chern Character, Connes-Chern Character, Family Index Theory, Topological Anomaly, BRST.
\end{abstract}

 \maketitle


\section{Introduction}\label{introduction}

\par Noncommutative geometry is one of the most prominent topics in theoretical physics. Through with last three decades it was extensively believed that the fundamental forces of the nature could be interpreted with more success via the machinery of noncommutative geometry and its different viewpoints \cite{maeda, connes-book2, connes-ch, lizzi}. Moyal noncommutative fields, inspired by fascinating formulations of strings, were proposed as the emergence of this idea in order to ramify the singular behaviors of quantum field theories, especially the quantum gravity.\footnote{See \cite{seiberg-witten} for a proper overview to this topic.} Appearing $\mathrm{UV/IR}$ mixing as a pathological feature of the Moyal quantum fields led to a concrete generalization of Moyal product as general translation-invariant noncommutative star products.\footnote{See \cite{lizzi-galluccio} and the references therein for a complete list of references and a short history on this topic.}

\par However, the topology and the geometry of noncommutative field theories with general translation-invariant star products have not been studied thoroughly yet. Actually, on the one hand it is a problem in noncommutative geometry, and on the other hand it is a physical behavior correlated to the topology and the geometry of the underlying spacetime and corresponding fiber bundles. But, in contrast to commutative algebras the correlation of noncommutative geometric machineries, such as cyclic (co)homology, and those of the ordinary differential structures of commutative geometry, such as the underlying spacetime, is not clear for noncommutative algebras. The former demands a point-free formulation, while the later one is based on ordinary topology and differentiation.

\par In this article we try to fill this gap by introducing a new cohomology theory, so called $\star$-cohomology, which can play an intermediate role between noncommutative and commutative differential geometry for noncommutative algebras. In the following we give a brief introduction to basic algebraic structures of $\star$-cohomology and in next sections we will develop it to study and formulate the topology of general translation-invariant noncommutative Yang-Mills theories.

\par Suppose that $X=Y\times Z$ is a $2n$-dimensional closed spin manifold with $\text{dim}Y=D$ and $Z=\mathbb{T}^{2m}$. Assume that there is a dense contractible open set, say $U \subset X$, which defines a chart with coordinate functions $x^\mu$, $\mu=0,\cdots, 2n-1$. Let $x^\mu$s split $X$, that is $x^\mu$ belongs to $Y$ for $\mu=0, \cdots, D-1$, and $x^i$ is a torus canonical coordinate for $D\leq i \leq 2n-1$. Here $x^0=t$ is the time parameter which varies from $-\infty$ to $+\infty$ on $U$. We also put a metric on $X$ unit diagonal for $x^\mu$s. Now consider a general translation-invariant noncommutative star product on $Z$, say $\star$, defined for 2-cocycle $\alpha$ as \cite {lizzi-galluccio, galluccio};
\begin{equation} \label {star}
(f \star g)(x)=\sum_{p,q \in \mathbb{Z}^{2m}} \tilde f(p)\tilde g(q)~e^{\alpha(p+q,p)}~e^{i(p+q).x/R}~,~~~~~f,g\in C^{\infty}(Z)~,
\end{equation}

\noindent for Fourier transformation
\begin{equation} \label {fourier}
\tilde f(p)=\frac{{1}}{{(2\pi R)^{2m}}}\int_Z f(x)~e_{-p}(x)~, ~~~~~f(x)=\sum_{p\in \mathbb{Z}^{2m}} \tilde f(p)~e_p(x)~,
\end{equation}

\noindent with the Fourier basis $e_p(x)=e^{ip.x/R}$, $p\in \mathbb{Z}^{2m}$. The integration in (\ref{fourier}) is taken for the Riemannian volume form of $Z$. Due to Hodge decomposition theorem in $\alpha^*$-cohomology \cite{varshovi1, varshovi2} one readily finds $\alpha=\alpha_M+\partial \beta$, where $\alpha_M$ is a Moyal 2-cocycle providing a Moyal star product $\star_M$, and $\beta$ is a 1-cocycle ($\beta(0)=0$ and $\overline{\beta}(p)=\beta(-p)$) with $\partial \beta(p,q)=\beta(p-q)-\beta(p)+\beta(q)$. In fact, $f' \star_M g'=(f \star g)'$ where
\begin{equation} \label {'}
f'(x)=\sum_{p\in \mathbb{Z}^{2m}}\tilde f(p)~e^{\beta(p)}~e_p(x)~,~~~~~f\in C^{\infty}(Z)~.
\end{equation}

\noindent Assume $H$ is a separable Hilbert space with coordinate function with orthonormal basis $\{ \left| p \right\rangle  \}_{p \in \mathbb{Z}^{2m}}$. Then, it can be checked that $\pi:C^{\infty}(Z) \to \mathcal{L}(H)$ with
\begin{equation} \label {rep}
\pi (f):=\hat f:=\sum_{p\in \mathbb{Z}^{2m}} \tilde f' (p)~\hat e_p~,~~~~~(\hat e_p)_{r,s}=e^{\alpha_M(r,s)}~\delta_{p,r-s}~,
\end{equation}

\noindent gives rise to a representation of $C^{\infty}_\star(Z)$ on $H$, i.e. $\hat f.\hat g=\widehat {f \star g}$, $f,g \in C^{\infty}(Z)$. This representation can be similarly extended to smooth functions on $X$. In this case for any $f \in C^{\infty}(X)$ the mapped element $\hat f$ is an operator valued function on $Y$. Let $\frak{C}_0$ be the algebra generated with $\hat f$, $f \in C^{\infty}(X)$. Then $\frak{C}_0$ is isomorphic to $C^{\infty}_\star(X)$ via $\pi$. This leads to definition of noncommutative polynomial. If $P(x_1,\cdots, x_k)$ is a polynomial of probably noncommutative variables $x_1,\cdots, x_k$, the noncommutative version of $P$, denoted by $P_\star$ is defined as:
\begin{equation} \label {polynomial}
P_\star(f_1,\cdots,f_k)=\pi^{-1}(P(\hat f_1,\cdots, \hat f_k))~.
\end{equation}

\par In fact, $\frak{C}_0$ is a unital $*$-algebra for $1=\hat 1$ and $\hat f^*=\hat f^{\dag}=\hat {\overline f}$. The domain of $\pi$ can be simply extended to vector and matrix valued smooth functions on $X$ with $v=(v_i) \to \hat v=(\hat v_i)$ and $g=(g_{ij}) \to \hat g=(\hat g_{ij})$. Moreover, if $\{ t^a\}$ is a basis for Lie algebra $\frak{g}$, then $\frak{g}$-valued smooth function $f=f^a t^a$ is mapped to $\hat f=\hat f^a t^a$ via $\pi$.


\par
\section{Star product, $\star$-Cohomology and Chern Characters}
\setcounter{equation}{0}

\par The representation map $\pi$ can be simply defined for differential forms on $X$. Let us denote the image of $k$-forms by $\frak{C}_k$ and set $\frak{C}=\oplus_{k=0}^{2n} \frak{C}_k$. Actually, $\frak {C}_k$ is the space of operator valued $k$-forms on $Y$ with extra dimensions on $Z$. Moreover, $\frak{C}$ is obviously a graded algebra with operator and wedge product. We define an exterior derivative operator on $\frak{C}_0$ as;
\begin{equation} \label {d}
d\hat f=\widehat {\partial_\mu f}~dx^{\mu}~,
\end{equation}

\noindent and extend it to $\frak{C}$. Then, $(\frak{C},d)$ is a differential graded algebra. The corresponding cohomology of $(\frak{C},d)$ is referred to as $\star$-cohomology and we denote its groups with $H_\star^k(X,\mathbb{C})$, $0 \leq k \leq 2n$. Since $\pi$ is an isomorphism and $C^{\infty}(X)=C^{\infty}_\star(X)$ as sets we readily find that $\pi :\Omega_k(X) \to \frak{C}_k$ is a bijective map for any $k\geq 0$, where here $\Omega_k(X)$ is the space of differential $k$-forms on $X$. Thus any element of $\frak{C}$ is simply shown with an overal symbol of $\hat ~$. Moreover, we readily see that $\pi \circ d_X=d \circ \pi$ for $d_X$ the exterior differential operator on $X$. Therefore, the following diagram commutes
\begin{equation} \label {diagram}
\begin{array}{*{20}{c}}
 0 & \to &  \Omega_0(X) & \to & \Omega_1(X) & \to & \cdots & \to & \Omega_k(X) & \to & \Omega_{k+1}(X) & \to & \cdots \\
~ & ~ &    \pi \downarrow  & ~ &  \pi \downarrow  & ~ & \cdots & ~ & \pi \downarrow  & ~ &  \pi \downarrow  & ~ & \cdots \\
 0 & \to &  \frak{C}_0 & \to & \frak{C}_1 & \to & \cdots & \to & \frak{C}_k & \to & \frak{C}_{k+1} & \to & \cdots \\
\end{array}~,
\end{equation}

\noindent wherein the upper arrows stand for $d_X$ and the lower ones indicate $d$. This provides isomorphisms $\pi_*:H_{\mathrm{dR}}^k(X,\mathbb{C}) \to H_\star^k(X,\mathbb{C})$. We are mostly interested in integral group $H^{2n}_\star(X,\mathbb{Z})$. To define it an integration structure is put on $\frak{C}$ as
\begin{equation} \label {integral}
\int \hat \omega=\left\{ {\begin{array}{*{20}{c}}
   \int_X f~,~~~~~\hat \omega=\hat f ~d^{2n}x \in \frak{C}_{2n}  \\
   ~~~0~,~~~~~~~~~~~~\text{other wise}~~~~~  \\
\end{array}} \right.
\end{equation}

\noindent where the integral is taken for Riemann volume form of $X$. It is easily seen that (\ref {integral}) defines a graded closed trace over $(\frak{C},d)$ for $\int d\hat \omega=0$ and $\int \hat \omega_1.\hat \omega_2=(-1)^{|\hat \omega_1||\hat \omega_2|} \int \hat \omega_2.\hat \omega_1$.
The space of classes $[\hat \omega] \in H_{\mathrm{dR}}^{2n}(X,\mathbb{C})$ with $\int \hat \omega \in \mathbb{Z}$ is referred to as $2n^{th}$ integral $\star$-cohomology group and is denoted with $H^{2n}_\star(X,\mathbb{Z})$. It can be seen that $\pi_*:H_{\mathrm{dR}}^{2n}(X,\mathbb{Z}) \to H_\star^{2n}(X,\mathbb{Z})$ is also an isomorphism. Therefore, due to Chern-Weil theory $H_\star^{2n}(X,\mathbb{Z})$ is generated with the $\pi$ image of $n^{th}$ Chern character $\mathrm{ch}_n(E)$ for vector bundles $E \to X$.

\par The noncommutativity via $\pi$ lets us to define three types of Chern characters. The first type Chern character is simply $\hat {\mathrm{ch}}_n(E)=\pi(\mathrm{ch}_n(E))$, which as mentioned above represents integral class in $H_\star^{2n}(X,\mathbb{Z})$. Let us denote it with $\hat{\mathrm{ch}}^{(1)}_n(E)$ to stress on its type. Second type Chern character is\footnote{From now on the notation $\mathrm{Tr}$ is used for trace on vector bundle dimensions or gauge group colors.}
\begin{equation} \label {second type}
\hat {\mathrm{ch}}_n^{(2)}(E)=\frac{{1}}{{(4\pi)^n n!}}~\mathrm{Tr}\left\{\hat F^n\right\}
\end{equation}

\noindent wherein $\nabla^2=- \frac{{i}}{{2}}F$ is the curvature of connection $\nabla$ on $\mathbb{C}^k \to E \to X$ and $\mathrm{Tr}$ is over $\mathbb{C}^k$. Hence;
$$\hat {\mathrm{ch}}_n^{(2)}(E)=\frac{{1}}{{(4\pi)^n n!}}~\epsilon^{\mu_1 \nu_1 \cdots \mu_n \nu_n}~\pi(\mathrm{Tr}\{ F_{\mu_1\nu_1}\star \cdots \star F_{\mu_n \nu_n} \})~d^{2n}x~.$$
\noindent It is obvious that if $n=1$ then $\hat {\mathrm{ch}}_n^{(2)}(E)$ and $\hat {\mathrm{ch}}_n^{(1)}(E)$ coincide and hence the cohomology class of $\hat {\mathrm{ch}}_n^{(2)}(E)$ is independent of connection $\nabla$. But, however, it is not the case in general;\footnote{Similar results for Moyal product are worked out in \cite {varshovi-dR-moyal}.}\\

\par \textbf{Theorem 1;} \emph{Suppose $\star$ is a general translation-invariant noncommutative star product on $X$ with $n \geq 2$. Then, the cohomology class of $\hat {\mathrm{ch}}^{(2)}_n(E)$ is independent of connection $\nabla$ if and only if $\star$ is a Moyal star product. Hence after;}
\par \textbf{a)} \emph{$\hat{\mathrm{ch}}^{(2)}_n(E)$ defines an integral cohomology class in $H_\star^{2n}(X,\mathbb{Z})$.}
\par \textbf{b)} \emph{$\hat{\mathrm{ch}}^{(2)}_n(E)$ and $\hat{\mathrm{ch}}^{(1)}_n(E)$ are cohomologous. That is;}
\begin{equation} \label {the1}
\int \hat{\mathrm{ch}}^{(2)}_n(E)=\int \hat{\mathrm{ch}}^{(1)}_n(E)=\int_X \mathrm{ch}_n(E) \in \mathbb{Z}~.
\end{equation}
~
\par \textbf{Hint to the Proof;} Actually, according to Chern-Weil theory it is enough to prove equality (\ref {the1}). The integral on the far left consists of the following integration according to (\ref {'});
$$\int_X  \epsilon^{\mu_1 \nu_1 \cdots \mu_n \nu_n}~\mathrm{Tr}\left\{ F'_{\mu_1\nu_1}\star_M \cdots \star_M F'_{\mu_n \nu_n} \right\}~,$$
\noindent where $\star_M$ is the Moyal product cohomologous to $\star$ via Hodge decomposition \cite {varshovi1}. It is also equal to
$$\int_X  \epsilon^{\mu_1 \nu_1 \cdots \mu_n \nu_n}~\mathrm{Tr}\left\{ F'_{\mu_1\nu_1} \cdots F'_{\mu_n \nu_n} \right\}~,$$
\noindent since Moyal product can be replaced by the ordinary product for integration of symmetric polynomials \cite{varshovi-dR-moyal}. This integral is independent of connection if and only if there exists some $C\in \mathbb{C}$ so that;
$$\int_X  \epsilon^{\mu_1 \nu_1 \cdots \mu_n \nu_n}~\mathrm{Tr}\left\{ F'_{\mu_1\nu_1} \cdots F'_{\mu_n \nu_n} \right\}=C\int_X  \epsilon^{\mu_1 \nu_1 \cdots \mu_n \nu_n}~\mathrm{Tr}\left\{ F_{\mu_1\nu_1} \cdots F_{\mu_n \nu_n} \right\}~.$$
\noindent But it is easily seen that the above equation holds if and only if 1-cocycle $\beta$ is linear and $C=1$. That is $\star=\star_M$. The rest of the proof is due to Chern-Weil theory. This finishes the theorem. \textbf{Q.E.D}\\

\par To define the third type Chern character we represent vector bundle $\mathbb{C}^k \to E\to X$ by $\pi$. Set
\begin{equation} \label {E}
\hat E=\{\hat V;~V\in C^{\infty}(E)\}~.
\end{equation}

\noindent Then, the connection $\nabla=d_X+A$, for $A \in \Omega_1(X) \otimes \mathbb{M}_k(\mathbb{C})$, is mapped to $\hat \nabla=d+\hat A$ on $\hat E$. Then, its curvature is an element of $\frak{C}_2 \otimes \mathbb{M}_k(\mathbb{C})$ and is equal to $\hat \nabla^2=d\hat A+\hat A^2=-\frac{{i}}{{2}}\hat F_\star$. The third type Chern character is then defined with $\hat \nabla^2$ as;
\begin{equation} \label {third type}
\hat {\mathrm{ch}}_n^{(3)}(E)=\frac{{1}}{{(4\pi)^n n!}}~\mathrm{Tr}\left\{\hat F_\star^n\right\}=\frac{{1}}{{(4\pi)^n n!}}~\epsilon^{\mu_1 \nu_1 \cdots \mu_n \nu_n}~\pi\left(\mathrm{Tr}\left\{ F_{\star \mu_1\nu_1}\star \cdots \star F_{\star \mu_n \nu_n} \right\}\right)~d^{2n}x~.
\end{equation}

\noindent We show $\hat {\mathrm{ch}}^{(3)}_n(E)$ with $\check{\mathrm{ch}}_n(E)$ for simplicity. The next theorem is significant to our formalism.\\

\par \textbf{Theorem 2;} \emph{For any general translation-invariant noncommutative star product $\star$ on $X$ we have;}
\par \textbf{a)} \emph{The cohomology class of $\check{\mathrm{ch}}_n(E)$ in $H^{2n}_\star(X,\mathbb{C})$ is independent of connection $\nabla$ ($\hat \nabla$).}
\par \textbf{b)} \emph{$\check{\mathrm{ch}}_n(E)$ represents an integral cohomology class in $H_\star^{2n}(X,\mathbb{Z})$.}
\par \textbf{c)} \emph{$\check{\mathrm{ch}}_n(E)$ and $\hat{\mathrm{ch}}^{(1)}_n(E)$ are cohomologous. That is;}
\begin{equation} \label {the2}
\int \check{\mathrm{ch}}_n(E)=\int \hat{\mathrm{ch}}^{(1)}_n(E)=\int_X \mathrm{ch}_n(E) \in \mathbb{Z}~.
\end{equation}
~
\par \textbf{Hint to the Proof;} For connection $\nabla=d_X+A$ we obtain;
$$\int \check{\mathrm{ch}}_n(E,A;\star)=\int \check{\mathrm{ch}}_n(E,A';\star_M)=\int_X \mathrm{ch}_n(E,A')=\int_X \mathrm{ch}_n(E)~,$$
\noindent where the first equality is due to (\ref {'}), the second one is the replacement of $\star_M$ with the ordinary product for integration of symmetric polynomials, and the last equation is an immediate consequence of Chern-Weil theory. This proves (\ref {the2}) and thus the theorem follows. \textbf{Q.E.D}


\par
\section{Index Theorem in $\star$-Cohomology and Abelian Anomaly}
\setcounter{equation}{0}

\par Assume vector bundle $\mathbb{C}^N \to E \to X$ with structure group $\mathrm{U}(N)$ for some $N$, in fundamental representation. The generators of $\frak{u}(N)$, say $t^a$s, are supposed to be Hermitian with totally anti-symmetric structure group $if^{abc}$, anti-commutator $\{t^a,t^b\}=-c^{abc} t^c$ and normalization condition $\mathrm{Tr}(t^at^b)=\frac{{1}}{{2}}\delta^{ab}$. We show the space of $\frak{g}$-valued differential forms on $X$ by $\tilde{\Omega}(X)=\oplus_{k=0}^{2n}~\tilde{\Omega}_k(X)$, where $\frak{g}=\frak{u}(N)$. Similarly the $\frak{C}_0 \otimes \frak{g}$-valued differential forms on $Y$ with extra dimensions on $Z$ are denoted as $\tilde {\frak{C}}$. We have the natural grading for $\tilde {\frak{C}}$ as $\tilde {\frak{C}}=\oplus_{k=0}^{2n}\tilde{\frak{C}}$ which for $d$ is defined accordingly. Thus, $(\tilde {\frak{C}},d)$ is a differential graded algebra on which $\mathrm{Tr} \otimes \int$ defines an integration structure or a closed graded trace. It is easy to see that $\tilde{\frak{C}}_0$ is a unital $*$-algebra with $1=\hat 1 \mathbb{I}$ and $(\hat f ^at^a)^*=\hat {f^a}^{\dag} {t^a}^{\dag}=\hat {\overline {f^a}} t^a$. We refer to it with $\cal{A}$. 

\par The involution $*$ is naturally extended to $\tilde{\frak{C}}_k$ for any $k \geq 0$. Conventionally, $\sigma \in \tilde{\frak{C}}$ is said to be Hermitian if $\sigma^*=\sigma$, and it is anti-Hermitian when $\sigma^*=-\sigma$. The set of anti-Hermitian elements of $\mathcal{A}$, denoted by $\tilde{\frak{g}}$, is in fact a Lie algebra. Each element of $\tilde{\frak{g}}$ also is known as infinitesimal gauge transformation. The Lie group generated by exponential of elements of $\tilde{\frak{g}}$, indicated with symbol $\tilde{G}$, is the gauge transformation group. Accordingly, the space of anti-Hermitian elements in $\tilde{\frak{C}}_1$, shown with $\Gamma$, is called the connection space. Any element of $\Gamma$, say $\hat A=-i\hat A^a_\mu t^a~dx^\mu$, for real functions $A^a_\mu \in C^{\infty}(X)$, is a connection form on mapped vector bundle $\hat E$ for connection $\hat \nabla=d+\hat A$ due to (\ref {E}).

\par The gauge transformation group $\tilde G$ (resp. $\tilde{\frak{g}}$) acts on $\Gamma$ form right:
\begin{equation} \label {action}
A  \triangleleft g=g^{-1}dg+g^{-1} A g~,~~~A \in \Gamma,~g\in \tilde G~~(\text{resp.~}A\triangleleft \alpha=d\alpha+[A,\alpha]~,~~~\alpha \in \tilde{\frak{g}}~)~.
\end{equation}

\par The Yang-Mills theory of vector bundle $SE:=S(X)\otimes E \to X$ is mapped to that of $\widehat {SE}$ which by definition is equipped with connection $\hat \nabla=d-i\hat A^a_\mu t^a~dx^\mu$ and curvature $\hat \nabla^2=-\frac{{i}}{{2}} \hat F_{\star \mu \nu}^a t^a~dx^{\mu} dx^{\nu}$. The Lagrangian and the action of the $\mathrm{U}(N)$-Yang-Mills theory is then given by;\footnote{For $\mathrm{U}(1)$ gauge theory one needs an overal factor of $\frac{{1}}{{2}}$. This is mandatory to compensate the normalization condition $\mathrm{Tr}(t^at^b)=\frac{{1}}{{2}} \delta^{ab}$ for $N >1$.}
\begin{equation} \label {y-m lagrangian}
L_{\mathrm{Y-M}}(\widehat{SE}, \hat{\nabla})=\mathrm{Tr} \left\{\hat \nabla^2.*\hat \nabla^2 \right\} \in \frak{C}_{2n}~,~~~~~S_{\mathrm{Y-M}}=\int L_{\mathrm{Y-M}}(\widehat{SE},\hat \nabla)~,
\end{equation}

\noindent for $*$ the Hodge star. Vector bundle $\widehat{SE}$, the space of spinors, is subject to Dirac operator $\mathcal D_0$ as
\begin{equation} \label {free dirac}
\mathcal{D}_0\hat \psi=i\gamma^{\mu}\partial_{\mu}\hat \psi~,
\end{equation}

\noindent on $U$ for Dirac matrices $\gamma^\mu$. Dirac operator $\mathcal D_0$ is usually perturbed to $\mathcal{D}_A$, for $A =-i \hat A^a_\mu t^a~dx^\mu \in \Gamma$;
\begin{equation} \label {non-free dirac}
\mathcal{D}_A\hat \psi=i\gamma^{\mu}\partial_{\mu}\hat \psi + \gamma^\mu t^a \hat A^a_\mu \hat \psi ~.
\end{equation}

\noindent The relevant Lagrangian and action are respectively
\begin{equation}
\label {dirac lagrangian}
L_{\mathcal{D}_0}(\widehat{SE})=\overline{\hat \psi}.\mathcal{D}_0\hat \psi~d^{2n}x=\hat \psi^{\dag} \gamma^0\mathcal{D}_0\hat \psi~d^{2n}x \in \frak{C}_{2n}~,~~~~~S_{\mathcal{D}_0}=\int L_{\mathcal{D}_0}(\widehat{SE})~,
\end{equation}
\begin{equation} \label {interaction lagrangian}
L_{\mathrm{int}}(\widehat{SE},\hat \nabla)=\overline{\hat \psi} \gamma^\mu  \hat A^a_\mu t^a \hat \psi ~d^{2n}x \in \frak{C}_{2n}~,~~~~~~~~~~~~S_{\mathrm{int}}=\int L_{\mathrm{int}}(\widehat{SE},\hat \nabla)~.
\end{equation}

\noindent The total action then defines the well-known noncommutative $\mathrm{U}(N)$-Yang-Mills theory;
\begin{equation} \label {total action}
S_{\mathrm{total}}=S_{\mathrm{Y-M}}+S_{\mathcal{D}_0}+S_{\mathrm{int}}=-\frac{{1}}{{4}}\int_X F_{\star \mu \nu}F_\star^{\mu \nu}+i\int_X \overline{\psi}\gamma^\mu\partial_\mu \psi +\int_X j^a_\star A^a_\mu~,
\end{equation}

\noindent for $j^{a\mu}_\star=\psi_{\beta,j}\star \overline{\psi}_{\alpha,i}~\gamma^{\mu}_{ij}~t^a_{\alpha \beta}$. Replacing $t^a$ and $\gamma^\mu$ respectively with $\mathbb{I}$ and $\gamma^\mu P_{\pm}$, for $P_{\pm}=\frac{{1 \pm \gamma }}{{2}}$ and
\begin{equation} \label {gamma5}
\gamma=\gamma_{2n+1}:=i^{n-1}\gamma^0 \cdots \gamma^{2n-1}~,
\end{equation}

\noindent leads to chiral singlet currents $j^{\mu}_{\star \pm}$, which obey the classical equation
\begin{equation} \label {conservation law}
\partial_\mu j^\mu_{\star \pm}+i[A^a_\mu,j^{a \mu}_{\star \pm}]_\star=0~.
\end{equation}

\noindent Let this equation be anomalously broken at quantum levels with appearing $\mathcal{A}_{\star \pm}$ on the right hand side. Then, since any translation-invariant noncommutative star product $\star$ is cyclic under integration the corresponding charges $Q_{\star \pm}(t)$ receive variations from $t=-\infty$ to $t=+\infty$ as;
\begin{equation} \label {delta q}
\Delta Q_{\star \pm}=Q_{\star \pm}(+\infty)-Q_{\star \pm}(-\infty)=\int_{-\infty}^{+\infty} \frac{{d}}{{dt}} Q_{\star \pm}(t)=\int_X \mathcal{A}_{\star \pm} =\int \hat {\mathcal{A}}_{\star \pm}~.
\end{equation}

\noindent Now let us make an \emph{ansatz} here. By this ansatz, so called \emph{physical consistency}, we assume that despite to appearence of consistent anomaly $\mathcal{A}_{\star \pm}$ the charge variation $\Delta Q_{\star \pm}$ respects the theory so that it is equal to an integer times the unit charge of it. Actually, the results achieved through with various methods of anomaly derivation in Moyal noncommutative gauge theories, such as noncommutative calculus \cite{langmann}, perturbative loop calculations \cite{ardalan, bonora, bonora2}, Seiberg-Witten map \cite{brandt} and matrix model \cite{varshovi-consistent},\footnote{We should emphasize that the results of \cite{varshovi-consistent} is in fact for general translation-invariant star products. See also \cite{varshovi-dR-moyal} and the references therein for a more complete list of such papers.} confirm the reasonability of the ansatz for the Moyal case. However, the idea for generalizing the results derived for Moyal product to general translation-invariant star products comes in principal from quantum equivalence theorem introduced in \cite {varshovi1, varshovi2} which asserts that the whole quantum behaviors of a noncommutative quantum field theory with an arbitrary translation-invariant star product coincide precisely with those of its Moyal product case of the same $\alpha^*$-cohomology class.

\par Thus, according to physical consistency ansatz, we demand the results of integrals in (\ref {delta q}) to be integers for any given translation-invariant noncommutative star product $\star$. Hence, for a homotopy of 2-cocyles as $s\alpha$ with $s\in [0,1]$ and correponding star product $\star_s$ we obtain $\int_X \mathcal{A}_{\star_s \pm} \in \mathbb{Z}$, which leads to $\frac{{d}}{{ds}}\int \int_X \mathcal{A}_{\star_s \pm}=0$ due to continuity. Furthermore,  for the commutative fields, i.e. $s=0$, we have; $\mathcal{A}_{\pm}=\mp \mathrm{ch}_n(SE)$. Therefore, $\hat {\mathcal{A}}_{\star \pm}$ and $\mp \hat{\mathrm{ch}}_n(SE)$ must be cohomologous in $H^{2n}_\star(X,\mathbb{Z})$. On the other hand, the left hand side of (\ref {conservation law}) transforms covariantly under infinitesimal gauge transformations via (\ref {action}) and $\hat \psi \to \alpha \triangleright \hat \psi:=- i \hat \alpha^a.t^a\psi$ for $\alpha =-i \hat \alpha^a~t^a \in \tilde{\frak{g}}$. Hence, $\hat {\mathcal{A}}_{\star \pm}$ is an equivariant form. That is, $\hat \omega=\hat {\mathcal{A}}_{\star \pm}\pm \check{\mathrm{ch}}_n(SE)$ is an equivariant form and thus it vanishes on $U$ for triviality of $E$ over it. Since $\hat \omega$ represents the null integral cohomology class in $H^{2n}_\star(X,\mathbb{Z})$ we readily conclude $\hat \omega=0$ . Thus, we have already established the following theoerm. \\

\par \textbf{Theorem 3;} \emph{The chiral Abelian anomaly in noncommutative Yang-Mills theories with general translation-invariant noncommutative star product $\star$ is given by $\mathcal{A}_{\star \pm}=\mp \check{\mathrm{ch}}_{n\star}(SE)$.}\\

\par It is well-known that the Dirac operator in the ordinary commutative case, i.e. $D_A=i\gamma^\mu\nabla_\mu$, $\nabla=d_X+A$, via $\mathbb{Z}_2$-grading of $SE \to X$ due to $\gamma$ of (\ref {gamma5}), is given as
\begin{equation} \label {dirac and gamma-commutative}
\gamma=\left( {\begin{array}{*{20}{c}}
   1 & 0  \\
   0 & -1  \\
\end{array}} \right)~,~~~~~D_A=\left( {\begin{array}{*{20}{c}}
   0 & D_A^-  \\
   D_A^+ & 0  \\
\end{array}} \right)~,~~~~~\gamma D_A=-D_A \gamma~,~~~~~{D_A^+}^{\dag}=D_A^-~.
\end{equation}

\par Following the main approach of Atiyah-Singer index theorem \cite{atiyah-dirac}\footnote{See also the approach of \cite {getzler}.} for $\star$-cohomology and also employing \textbf{Theorem 3} we readily find an index formula for any translation-invariant noncommutative anomalies via the machinery of $\star$-cohomology. Show $\pi(D_A^{(\mp)})$ with $\mathcal{D}_A^{(\mp)}$. \\

\par \textbf{Theorem 4;} \emph{For any noncommutative $\mathrm{U}(N)$-Yang-Mills theory with general translation-invariant noncommutative star product $\star$ the topological index of $\mathcal{D}_A^{\mp}$ is given by $\check{\mathrm{ch}}_n(SE)$ as;}
\begin{equation} \label {theorem4}
\int_X \mathcal{A}_{\star \pm}=\mp \int \check{\mathrm{ch}}_n(SE)=\emph{\emph{Index}}(\mathcal{D}_A^{\mp})~.
\end{equation}
~
\noindent The topological index is given for de Rham integral class in $H_{\mathrm{dR}}^{2n}(X,\mathbb{Z})$ with $\check{\mathrm{ch}}_{n \star}(SE)$ via (\ref {polynomial});
$$\emph{\emph{Index}}(\mathcal{D}_A^{\mp})=\int_X \mathcal{A}_{\star \pm}=\mp \int_X \check{\mathrm{ch}}_{n \star}(SE)~.$$


\par
\section{Anomalies, $\star$-Cohomology and the Connes-Chern Characters}
\setcounter{equation}{0}

\par In previous section we established an intimate correlation between $\star$- and de Rham cohomology to figure out the topological structure in the background of a translation-invariant noncommutative Yang-Mills theory. In this section we try to demonstrate a similar relation for cyclic (co)homology and the corresponding Connes-Chern character.\footnote{See \cite{varshovi-dR-moyal} for the special case of Moyal star product.} One should remember that since $\mathcal{A}$ (or $C^{\infty}_\star(X)$) is a noncommutative algebra there is no definite coincidence for de Rham and cyclic (co)homology (in the sense of \cite {connes-book} via Hochschild-Kostant-Rosenberg formula\footnote{See also \cite {connes-differential, khalkhali}.}). Therefore, topological interpretation of noncommutative anomalies need some breakthrough between cyclic (co)homology of $\mathcal{A}$ on the one hand and de Rham cohomology of the topologically commutative underlying spacetime $X$ on the other hand. In this section we proved that this subjective is properly achieved by applying the machinery of $\star$-cohomology.

Consider a trivial vector bundle over $X$, say $\mathbb{C}^{N'} \to E'=\mathbb{C}^{N'} \times X \to X$, for some large enough $N'$, so that $E \to X$ is embedded in via the image of an idemponent $e \in \mathbb{M}_{N'}(C^{\infty}(X))$, i.e.;
$$C^{\infty}(E)=\{\sigma \in C^{\infty}(E');~e.\sigma =\sigma \}~.$$
\noindent Then, ${SE}':=S(X)\otimes E' \to X$ is subject to Dirac operator $D=i\gamma^\mu \partial_\mu$ on $U$. Actually, $\widehat{SE}'$ could be completed to a Hilbert space, $\cal H$, with ordinary inner product $\left\langle {\rho(\psi_1)} \mathrel{\left | {\vphantom {\rho(\psi_1) \rho(\psi_2)}} \right. \kern-\nulldelimiterspace} {\rho(\psi_2)} \right\rangle =\int_X\overline{\psi}_1\psi_2$. We see that $\pi$ and the Dirac operator $D$ commute, i.e. $\pi \circ D=\mathcal{D} \circ \pi$ for $\mathcal{D}=i\gamma^\mu\partial_\mu$, and therefore, the spectrum of $\mathcal{D}$ coincides with that of $D$. Hence, Dirac operator $\mathcal{D}$ is a densely defined unbounded Hermitian operator on $\mathcal{H}$. Thus $(\mathcal{A},\mathcal{H},\mathcal{D})$ is a spectral triple. The action of unital $*$-algebra $\cal{A}$ on $\cal H$ is also given as $( \hat \alpha^a t^a)\triangleright \hat \psi=\hat \alpha^a. \hat t^a \psi$. In addition, we have a $\mathbb{Z}_2$-grading of $\mathcal{H}$ as $\mathcal{H}=\mathcal{H}^+\oplus \mathcal{H}^-$ with;
\begin{equation} \label {dirac and gamma}
\mathcal{D}=\left( {\begin{array}{*{20}{c}}
   0 & \mathcal{D}^-  \\
   \mathcal{D}^+ & 0  \\
\end{array}} \right)~,~~~~~\gamma \mathcal{D}=-\mathcal{D} \gamma~,~~~~~{\mathcal{D}^+}^{\dag}=\mathcal{D}^-~,
\end{equation}

\noindent for $\gamma$ of (\ref {dirac and gamma-commutative}). It is also clear that $\gamma \alpha=\alpha \gamma$, $\alpha \in \mathcal{A}$. Now we employ the homotopy of last section for star rpoducts $\star_s$. Then, it can be seen that for any $\hat a \in \mathcal{A}$, the operator $[\mathcal{D},\hat a]$ is homotopic to $[D,a]$ and thus is a densely defined operator which could be extended to a bounded operator on $\mathcal{H}$. Also for any $p > 2n$ we have $(1+\mathcal{D}^2)^{-1} \in \mathcal{L}^{p/2}(\mathcal{H})$ since $\emph{\emph{spec}}(\mathcal{D})=\emph{\emph{spec}}(D)$ \cite {gilky}. Set $F=\mathcal{D}/|\mathcal{D}|$. Actually, $F$ is a bounded operator with $\gamma F=-F \gamma$ and $F^2=1$. Hence, $(\mathcal{H},F,\gamma)$ is an even $p$-summable Fredholm module over $\mathcal{A}$, and an element of $K$-homology group $K^0(\mathcal{A})$.

\par The vector bundle $\widehat{SE}$ is in fact a dual element with respect to $(\mathcal{H},F,\gamma)$ in $K$-theory group $K_0(\mathcal{A})$. Assume an idemponent $e \in \mathbb{M}_{N'}(\mathcal{A})$ so that $\widehat{SE}$ embeds in $\widehat{SE}'$ as $\widehat{SE}=\{\hat \psi \in \widehat{SE}';~e\hat \psi=\hat \psi \}$. The connection on $\widehat{SE}$ is canonically defined with $\hat{\nabla} \hat\psi=e. d\hat \psi$. However, it is seen that if for a local basis over $U$ we define $\hat{\nabla}=d+A=d-i\hat A^a_\mu t^a~dx^\mu$, then; $\emph{\emph{Index}}(F^+_e)=\emph{\emph{Index}}(\mathcal{D}_A^+)$, in which
$$F=\left( {\begin{array}{*{20}{c}}
   0 & F^-  \\
   F^+ & 0  \\
\end{array}} \right)~,
$$
\noindent for the $\mathbb{Z}_2$-grading of $\gamma$ and $F^+_e=eF^+e:e\mathcal{H}^+ \to e\mathcal{H}^-$. Hence, according to noncommutative index theorem and \textbf{Theorem 4} we find;\\

\par \textbf{Theorem 5;} \emph{The integral cohomology class of the third type Chern character $\check{\mathrm{ch}}_n(SE)$ in $H_\star^{2n}(X,\mathbb{Z})$ (and the corresponding topological index of chiral Abelian anomaly) is given by the pairing of Connes-Chern characters of $2n^{th}$ cyclic (co)homology groups due to $K$-theory and $K$-cohomology. That is;}\footnote{The notation $\mathrm{Trace}$ is used for trace of operators on the corresponding Hilbert space of Dirac operator.}
\begin{equation} \label {1}
\int \check{\mathrm{ch}}_n(SE)=\left\langle \emph{\emph{Ch}}^{2n}(\mathcal{H},F,\gamma), \emph{\emph{Ch}}_{2n}[e]  \right\rangle =\frac{{(-1)^n}}{{2}} \mathrm{Trace}\{ \gamma F[F,e][F,e] \cdots [F,e] \} ~,
\end{equation}

\noindent \emph{for $2n+1$ copies of $e$ and for $\mathrm{Trace}$ the trace of operators.}\\

\par We remember that $\emph{\emph{Ch}}^{2n}(\mathcal{H},F,\gamma)$ in (\ref {1}) is the Connes-Chern character of $(\mathcal{H},F,\gamma) \in K^0(\mathcal{A})$ as
\begin{equation} \label {connes-chern}
\emph{\emph{Ch}}^{2n}(\mathcal{H},\mathcal{D},\gamma)(a_0,a_1,\cdots,a_{2n})=\left(-1\right)^n\left(\frac{{n!}}{{2}}\right) \mathrm{Trace}\left\{ \gamma F[F,a_0][F,a_1] \cdots [F,a_{2n}] \right\}~,
\end{equation}

\noindent for $a_0,a_1,\cdots,a_{2n} \in \mathcal{A}$, and
\begin{equation} \label {connes-chern-homology}
\emph{\emph{Ch}}_{2n}[e]= \sum_{k=0}^n (-1)^k\frac{{(2k)!}}{{k!}}~\mathrm{tr}\left\{ \left(e-\frac{{1}}{{2}}\right)\otimes e^{2k\otimes} \right\}~,
\end{equation}

\noindent wherein $e \in \mathbb{M}_{N'}(\mathcal{A})$ represents the $K$-theory class $[e]$ in $K_0(\mathcal{A})$ and $\mathrm{tr}$ is the trace of $\mathbb{C}^{N'}$.

\par As we mentioned above $\star$-cohomology plays an intermediate role between de Rham and cyclic (co)homologies for any general translation-invariant noncommutative star product $\star$. In previous sections we established an intimate correlation of $\star$- and de Rham cohomology theories. Now, by employing the special abilities of $\star$-cohomology, due to its partly commutative geometric structures, we can also prove a geometric correspondence between $\star$- and cyclic (co)homology. We emphasize that this relation must be implemented in the setting we just apply for demonstrating that of de Rham and cyclic (co)homologies for commutative algebras. By means of $\star$-cohomology we find the same arguments even for noncommutative algebra $\mathcal{A}$ (or $\frak{C}_0$).

\par This correlation is easy to see within familiar concepts of noncommutative geometry. Actually, the well-known Hochschild-Kostant-Rosenberg map $\alpha: HH_{k}(\mathcal{A}) \to \Omega_k(X)$, for $HH_{*}(\mathcal{A})$ the Hochschild homology group, leads to an isomorphism, similarly denoted by $\alpha$, between cyclic homology group $HC_{2n}(\mathcal{A})$ and $\oplus_{j=0}^{n}H_\star^{2j}(X,\mathbb{C})$ due to commutativity of $d \circ \pi =\pi \circ d_X$. It is not hard to see that $\alpha$ produces the third type Chern chracter $\check{\mathrm{ch}}_n(SE)$ from Connes-Chern character $\mathrm{Ch}_{2n}[e] \in HC_{2n}(\mathcal{A})$. In fact, the composition of isomorphism $\alpha: HC_{2n}(\mathcal{A}) \to \oplus_{j=0}^nH_{\star}^{2j}(X,\mathbb{C})$ and the canonical projection $\pi_{2n}:\oplus_{j=0}^nH^{2j}_{\star}(X,\mathbb{C}) \to H^{2n}_{\star}(X,\mathbb{C})$ leads to the following result;
\begin{equation} \label {lambda}
\Xi=\left(\frac{{1}}{{2i\pi }}\right)^n \frac{{1}}{{(2n)!}}~\pi_{2n}\circ \alpha: HC_{2n}(\mathcal{A}) \to H^{2n}_{\star}(X,\mathbb{C})~,~~~~~\Xi([\mathrm{Ch}_{2n}[e]])=[\check{\mathrm{ch}}_{n}(SE)]~,
\end{equation}

\noindent for corresponding cohomology classes $[\mathrm{Ch}_{2n}[e]]$ and $[\check{\mathrm{ch}}_{n}(SE)]$. The homomorphism $\Xi$ can be derived with more detailed formalism. To see this we note that for connection $\hat{\nabla}=e.d=d-i \hat A^a_\mu t^a~dx^\mu$ the curvature $\hat \nabla^2$ is actually $e.de.de=-\frac{{i}}{{2}}\hat F_{\star \mu \nu}~dx^\mu dx^\nu$. However, with $(e.de.de)^n=e.(de)^{2n}$ we readily find $\alpha\left(\mathrm{Ch}_{2n}[e]\right)=\left(2i\pi \right)^n (2n)!~\check{\mathrm{ch}}_n(SE)+\Delta$, where $\Delta$ is a direct summation of an exact $2n$-form and closed forms of lower even orders in $\frak{C}$. Now let $\Omega^*(\mathcal{A})= \oplus_{k\geq 0}\mathcal{A}^{(k+1)\otimes}$ and consider $\xi:\Omega^*(\mathcal{A})\to \frak{C}$, with $\xi=0$ on $\mathcal{A}^{(k+1)\otimes}$ for $k > 2n$, and
\begin{equation} \label {lambda'}
\xi \left( a_0\otimes \cdots \otimes a_{k}\right) =-\left(\frac{{-1}}{{2\pi }}\right)^{n} \frac{{i}}{{2^n(2n)!}}~{\mathrm{Tr}}\{\gamma a_0[\mathcal{D}, a_1]\cdots [\mathcal{D}, a_{k}]\}~d^{2n}x~,
\end{equation}

\noindent for $k \leq 2n$. Here $\mathrm{Tr}$ is the trace on Dirac matrices and on the colors. Thereby, we readily find that $\xi(\emph{\emph{Ch}}_{2n}[e])=\check{\mathrm{ch}}_{n}(SE)+d \phi$ for some $\phi \in \frak{C}_{2n-1}$. We have the following lemma. \\

\par \textbf{Lemma 1;} \emph{The linear map $\xi$ of (\ref {lambda'}) leads to a surjection from cyclic cohomology group $HC_{2n}(\mathcal{A})$ to $H^{2n}_{\star}(X,\mathbb{C})$, i.e. $\xi_*:HC_{2n}(\mathcal{A}) \to H^{2n}_{\star}(X,\mathbb{C})$ for any general translation-invariant noncommutative star product $\star$. Moreover, $\xi_*$ is in fact a redefinition for Hochschild-Kostant-Rosenberg map due to (\ref {lambda}) with $\xi_*= \Xi$.} \\

\par \textbf{Proof;} First note that $[\mathcal{D},a]=i\gamma^\mu \partial_\mu a$, $a\in \mathcal{A}$. Then since
$$\mathrm{Tr}\left(\gamma \gamma^{\mu_1}\cdots \gamma^{\mu_{2n}}\right)=(-i)^{n-1}2^n~\epsilon^{\mu_1 \cdots \mu_{2n}}~,~~~~~\mathrm{Tr}\left(\gamma \gamma^{\mu_1}\cdots \gamma^{\mu_{k}}\right)=0~,~~~~~k<2n~,$$
\noindent it is seen that $\xi$ vanishes on $\mathcal{A}^{(k+1)\otimes}$ for all $k \ne 2n$. On the other hand, $\phi:=\int \circ \xi$ is a closed graded trace on $\Omega(\mathcal{A})$ with support in $\mathcal{A}^{(2n+1)\otimes}$. Thus, it defines a cyclic cohomology class in $HC_{2n}(\mathcal{A})$. This proves that the Connes' $(b,B)$-bicomplex is compatible with the de Rham complex via $\xi$ so that $\xi \circ b=0$ and $\xi \circ B \in d \frak{C}$. Thus $\xi$ is reduced to a well-defined map $\xi_*:HC_{2n}(\mathcal{A}) \to H^{2n}_{\star}(X,\mathbb{C})$. Direct calculation shows that $\xi_*=\Xi$. This finishes the lemma. \textbf{Q.E.D} \\

\par Therefore, we have already established the following theorem.\\

\par \textbf{Theorem 6;} \emph{For any general translation-invariant noncommutative star product $\star$ on $X$ the integral cohomology class of the third type Chern character $\check{\mathrm{ch}}_n(SE)$ in $H_\star^{2n}(X,\mathbb{Z})$ is the image of that of $\emph{Ch}_{2n}[e]$ in cyclic cohomology group $HC_{2n}(\mathcal{A})$ via $\xi$ due to (\ref {lambda'}).}\\

\par \textbf{Corollary 1;} \emph{The topological index of chiral Abelian anomaly for any general translation-invariant noncommutative $\mathrm{U}(N)$-Yang-Mills theory is given by Connes-Chern character of the corresponding vector bundle $\widehat{SE}$. That is;}
$$\text{Index}(\mathcal{D}_A^{\mp})=\int \hat{\mathcal{A}}_{\star \pm}=\mp \int \xi \left(\emph{\emph{Ch}}_{2n}[e]\right)~.$$


\par
\section{Family Index and Homotopy Class of Topological Anomaly}
\setcounter{equation}{0}

\par Topological or consistent anomalies in noncommutative field theories have been considered by several authors \cite{langmann, bonora, brandt, varshovi-consistent},\footnote{In \cite{langmann, bonora, brandt} the authors considered the Moyal product and in \cite {varshovi-consistent} general translation-invariant noncommutative star product has been considered via some matrix model.} but however, the topological/geometric meaning of the solutions and of the corresponding formulations remained unclear especially for the case of general translation-invariant noncommutative star products. Thus, in this section the subjective is to studying topological anomalies of translation-invariant noncommutative Yang-Mills theories via homotopy classes due to Bismut-Freed determinant bundle and the family index approach \cite {bismut-freed1, bismut-freed2, varshovi-cyclic-moyal}.

\par Let us assume the action of $\tilde G$ on $\Gamma$ of (\ref {action}) is free so that $\Gamma \to \Gamma /\tilde G$ provides a principal $\tilde{G}$-bundle. We also suppose that $\tilde{X}:=\Gamma /\tilde G$ is a smooth manifold. Remember that for any $A \in \Gamma$, the Dirac operator is perturbed to unbounded Hermitian operator $\mathcal{D}_A:=\mathcal{D}+A$. Thus, $(\mathcal{H},\mathcal{D}_A,\gamma)$ is also regarded as an even $p$-summable Fredholm module homotopic to $(\mathcal{H},\mathcal{D},\gamma)$. Also
$$\mathcal{D}_A=\left( {\begin{array}{*{20}{c}}
   0 & \mathcal{D}_A^-  \\
    \mathcal{D}_A^+ & 0  \\
\end{array}} \right)~,$$
\noindent 

\noindent according to $\mathbb{Z}_2$-grading of $\gamma$. Thus, $\mathrm{Ker} (\mathcal{D}_A^+) \subset \mathcal{H}^+$ and $\mathrm{Ker} (\mathcal{D}_A^-) \subset \mathcal{H}^-$ are finite dimensional. Consider the Bismut-Freed determinant line bundle \cite{bismut-freed1, bismut-freed2}
\begin{equation} \label {determinant bundle}
\emph{\emph{Det}}:=\emph{\emph{Det}}(\mathcal{H},\mathcal{D},\gamma):=\emph{\emph{det}}(\mathrm{Ker} (\mathcal{D}_A^+))\otimes \emph{\emph{det}}(\mathrm{Ker} (\mathcal{D}_A^-))~,
\end{equation}

\noindent which has a natural metric and unitary connection, say $\nabla^{\emph{\emph{Det}}}=d_{\Gamma}-2\pi i \Pi(A)$, where $\Pi(A)$ is a one-form on $\Gamma$, and $A \in \Gamma$. Here, $d_\Gamma$ is the exterior differential operator on $\Gamma$. Actually $\Pi(A)$ is a closed form when it is restricted to orbits of $\tilde{G}$ \cite{bismut-freed1}. Therefore, the connection of $\nabla^{\emph{\emph{Det}}}$ is flat. That is;
\begin{align}
\label {slavnov-taylor}
2\pi i ~ \Pi(A)=\delta W(A)~,~~~~~\delta \Pi(A)=0~,
\end{align}

\noindent where $\delta$ is the exterior differential operator on $\tilde G$, the BRST operator. Due to Bismut-Freed results $\Pi$ represents a non-trivial cohomology class in $H^1_{\mathrm{dR}}(\tilde{G},\mathbb{C})$, since in fact $W(A)$, the quantum action, is not in general a smooth function on $\Gamma$. The BRST closedness of $\Pi$, the Wess-Zumino consistency condition, is then an immediate consequence of (\ref {slavnov-taylor}). Actually, $\frak{G}(A):=2\pi i~\Pi(A)$ is the topological anomaly. The flatness of $\nabla^{\emph{\emph{Det}}}$ on orbits of $\tilde G$ implies that the parallelism structure on $\emph{\emph{Det}}$ provides a covering space for Lie group $\tilde G$. Thus, if $\frak{i}:S^1 \to \tilde G$ defines a smooth map then;\footnote{Actually, for commutative $\mathrm{SU}(N)$-Yang-Mills theories $\mathbb{Q}$ can be replaced with $\mathbb{Z}$. But for the noncommutative case the gauge group $\mathrm{SU}(N)$ has to be replaced with $\mathrm{U}(N)$, since $F_\star$ and the Lagrangian contain anti-commutators of the corresponding Lie algebra elements. This, leads to a non-trivial fundamental group for the gauge transformation group $\tilde G$ \cite{hatcher}. Thus, despite to ordinary Yang-Mills theories, for those with noncommutative star products we should deal with rational cohomology $H^*_{\mathrm{dR}}(\tilde G,\mathbb{Q})$ instead of $H_{\mathrm{dR}}^*(\tilde G,\mathbb{Z})$ for topological anomalies. This is in fact due to $\mathrm{U}(1)$ component in $\mathrm{U}(N)=\mathrm{U}(1)\times \mathrm{SU}(N)$. The only finite covering of $\mathrm{U}(1)$ is $\mathrm{U}(1)$ itself. Therefore, the result of (\ref {deck}) belongs to $\frac{{1}}{{l}}\mathbb{Z}$ for $l$-covering. For more details see \cite{bertlman, varshovi-cyclic-moyal, griffith}.}
\begin{equation} \label {deck}
\int_{S^1}\frak{i}^*(\Pi) \in \mathbb{Q}~.
\end{equation}

\noindent Therefore, we readily find that $\Pi \in H^1_{\mathrm{dR}}(\tilde{G},\mathbb{Q})$. In principal, the equality of (\ref {deck}) should be invariant with respect to homotopy of 2-cocycles $s\alpha$ for $s\in [0,1]$ and its corresponding translation-invariant noncommutative star producs $\star_s$, with $\star_1=\star$ and $\star_0$ the ordinary product. To see this let $\tilde G_s$ be the gauge transformation group defined with $\star_s$. That is $\tilde G=\tilde G_1$ and $\tilde G_0$ is the gauge transformation group in commutative $\mathrm{U}(N)$-Yang-Mills theory. Then $M=\{\tilde G_s;~s\in I=[0,1]\}$ is topologically a cylindrical space with ends $\tilde G$ and $\tilde G_0$. Actually, $M$ is topologically equivalent to $I \times \tilde G$ and therefore it deformation retracts on both $\tilde G$ and $\tilde G_0$ \cite{hatcher, varshovi-cyclic-moyal}. Put a smooth structure on $M$ with $d_M=\delta +ds\otimes \partial/\partial s$. Note that here $\delta$ is the exterior derivative operator on $\tilde G_s$ for any $s\in [0,1]$. Now, $\Pi$ can be assumed as a one-form on $M$. By homotopy invariance of (\ref {deck}) we see that, there is some $\Phi \in C^{\infty}(M)$ so that $\frac{{d}}{{ds}}\Pi =\delta \Phi$. In other words, if $\Pi_0$ be the Bismut-Freed connection of commutative Yang-Mills theory;
\begin{equation} \label {pipi}
\Pi-\Pi_0=\delta\left( \int_0^1\Phi~ds\right)~.
\end{equation}

\noindent Moreover, $\tilde G$ and $\tilde G_0$ are homotopic equivalent. Let $f:\tilde G \to \tilde G_0$ defines this equivalence relation. Then (\ref {pipi}) leads to the following theorem. \\

\par \textbf{Theorem 7;} \emph{Let $\star$ be a general translation-invariant noncommutative star product on $X$. Also suppose that $[\frak{G}_0(A)]$ (resp. $[\frak{G}(A)]$) is the cohomology class of topological anomaly of commutative (resp. $\star$-noncommutative) $\mathrm{U}(N)$-Yang-Mills theory in $H^1_{\mathrm{dR}}(\tilde{G}_0,\mathbb{C})$ (resp. $H^1_{\mathrm{dR}}(\tilde{G},\mathbb{C})$). Then, we have; $f^*([\frak{G}_0(A)])=[\frak{G}(A)]$. Moreover, $\frak{G}_0(A)/2\pi i$ (resp. $\frak{G}(A)/2\pi i$) defines a rational cohomology class in $H^1_{\mathrm{dR}}(\tilde{G}_0,\mathbb{Q})$ (resp. $H^1_{\mathrm{dR}}(\tilde{G},\mathbb{Q})$).} \\

\par We should explain $f$ with more details. Let $\tilde{\frak{g}}_s$ be the infinitesimal gauge transformation group as the Lie algebra of $\tilde G_s$. Moreover, let $\pi_s$ be the corresponding representation map for star product $\star_s$. Therefore, $\{ \pi_s(e_p) t^a\}$ for $p \in \mathbb{Z}^{2m}$ and $a=0,\cdots , N^2-1$ provides a basis for $\tilde {\frak{g}}_s$. Set $df:\tilde{\frak{g}} \to \tilde{\frak{g}}_0$ with $df(\pi(e_p) t^a)=\pi_0(e_p) t^a$. Then, it is seen that its integral provides a group isomorphism $f:\tilde G \to \tilde G_0$.\footnote{For more details of this proof see \cite{hall}. In \cite{varshovi-cyclic-moyal} we also proved similar results for the special case of Moyal product.}  Actually, $f^{*}$ replaces the ordinary product with translation-invariant noncommutative star product $\star$ within differential forms. This leads us to the following corollary.\\

\par \textbf{Corollary 2;} \emph{Assume that the topological anomaly of commutative $\mathrm{U}(N)$-Yang-Mills gauge theory, $\frak{G}_0(A)$, is given by integration of polynomial $P(A,C)$ over spacetime $X$ as;
$\frak{G}_0(A)=\int_X P(A,C)$, for $C$ the ghost field. Then, the topological anomaly of noncommutative $\mathrm{U}(N)$-Yang-Mills theory for any translation-invariant noncommutative star product $\star$ is;}
\begin{equation} \label {corollary2}
\frak{G}(A)=\int_X P_\star(A,C) ~.
\end{equation}
\noindent \emph{wherein $P_\star$ is the noncommutative polynomial of $P$ due to (\ref {polynomial}).} \\ 

\par Note that according to \textbf{Theorem 7} we know that
$$\frak{G}(A)=\int_XP_\star(A,C)+\delta \Phi(A)~$$
\noindent for BRST exact term $\delta \Phi(A)$. Adding $-\Phi(A)$ to the quantum action $W(A)$ as a counter term, i.e. $W'(A)=W(A)-\Phi(A)$, to renormalize the theory accordingly, then we obtain the topological anomaly $\frak{G}(A)$ as described in (\ref {corollary2}). In principal, the topological anomaly of noncommutative $\mathrm{U}(N)$-Yang-Mills gauge theory on $4$-dimensional spacetime $X$ and for any translation-invariant noncommutative star product $\star$ is given for Weyl fermions as;\footnote{Here the notation $\mathrm{tr}$ is used for trace on both group colors and matrix representation due to $\pi$.}
\begin{equation} \label {corollary3}
\frak{G}(A)=\frac{{1}}{{24\pi^2}} \int \mathrm{tr}\{C\star d(A\star dA+\frac{{1}}{{2}}A^3) \}~.
\end{equation}

\par Up to now we studied the topological anomaly within homotopy classes in $H^1_{\mathrm{dR}}(\tilde G,\mathbb{Q})$ via family index theory due to Bismut-Freed determinant line bundle. However, it can also be regarded as a noncommutative geometric problem through with the machinery of cyclic (co)homology \cite {perrot, perrot2, perrot1}.

\par To see this we note that the inclusion $\check g_A:\tilde G \to \Gamma$ as an orbit which passes $A\in \Gamma$ is an invertible element of $C^{\infty}(\Gamma)\otimes \mathcal{A}$, and hence defines a class in $K_1^{\emph{\emph{alg}}}(C^{\infty}(\Gamma)\otimes \mathcal{A})$. Its Chern character in odd periodic cyclic homology $HP_{\mathrm{odd}}(C^{\infty}(\tilde{G}\otimes \mathcal{A})$ is in fact an element of differential graded algebra $(\Lambda',\hat d)$ with $\Lambda':=\oplus_{k\geq 0} \Lambda'_k ( C^{\infty}(\tilde G)\otimes \mathcal{A})$ and $\hat d=\delta+d$. For the Maurer-Cartan form $\omega_A=\check g_A^{-1}\hat d\check g_A$ we read
\begin{equation} \label {ch*}
\emph{\emph{Ch}}^1_*[\check g_A]=\sum_{k\geq 0}~(-1)^k \frac{{k!}}{{(2k+1)!}}\omega_A^{2k+1}~\in HP_{\mathrm{odd}}( C^{\infty}(\tilde G)\otimes \mathcal{A})~.
\end{equation}

\noindent for $\emph{\emph{Ch}}^1_*[\check g_A]$ the corresponding Connes-Chern character in odd periodic cyclic homology for $\check g_A$. An important formula is \cite {perrot, connes-moscovici-cs};
\begin{equation} \label{perrot}
CS(\mathcal{H},\mathcal{D},\gamma)=\left\langle {\emph{\emph{Ch}}_*^1[\check g_A],\emph{\emph{Ch}}_1^*(\mathcal{H},\mathcal{D},\gamma)} \right\rangle~,
\end{equation}

\noindent wherein $\emph{\emph{Ch}}_1^*(\mathcal{H},\mathcal{D},\gamma)$ is the class of Connes-Chern character in odd periodic cyclic cohomology $HP^{\mathrm{odd}}(C^{\infty}(\tilde G)\otimes \mathcal{A})$ via (\ref {connes-chern}) and $CS(\mathcal{H},\mathcal{D},\gamma)$ is the corresponding Chern-Simons form. The pairing (\ref {perrot}) is in fact taken place via cup product
$$\cup : H_k(\tilde G,\mathbb{C})\otimes HC^j(\mathcal{A}) \to HC^{k+j}(C^{\infty}(\tilde G)\otimes \mathcal{A})~,$$
\noindent and hence, it represents de Rham cohomology classes in $H_{\mathrm{dR}}^{\mathrm{odd}}(\tilde{G},\mathbb{C})$. The main result is that the component in $H_{\mathrm{dR}}^1(\tilde{G},\mathbb{C})$ of pairing (\ref {perrot}) coincides with $\Pi(A)$ up to some exact form on $\tilde G$. Therefore, if $i:S^1 \to \tilde G$ is a smooth map, then we find;
\begin{equation} \label {perrot-1} 
\int_{S^1}i^*(\frak{G}(A))=2\pi i \left\langle {\emph{\emph{Ch}}_*^1[\check g_A],[S^1]\cup \emph{\emph{Ch}}_2^*(\mathcal{H},\mathcal{D},\gamma)} \right\rangle\in 2\pi i ~\mathbb{Q}~,
\end{equation} 

\noindent where $i:S^1 \to \tilde G$ is considered to represent a class in $H_1(\tilde G,\mathbb{C})$, say $[S^1]$. Calculating (\ref {perrot-1}) is actually accomplished by using Connes-Moscovici local index formula as a residue of some zeta function.\footnote{See \cite {perrot} for more details.}


\par
\section{Summary and Conclusions}
\setcounter{equation}{0}

\par Through this paper we introduced a cohomology theory, so called $\star$-cohomology, with cohomology groups $H_\star^k(X,\mathbb{C})$, on spacetime manifold $X$, to describe general translation-invariant noncommutative quantum field theories by means of both commutative and noncommutative geometric structures. In fact, $\star$-cohomology plays an intermediate role between de Rham and cyclic (co)homology theories for noncommutative algebras. It provides a breakthrough between commutative and noncommutative quantum field theories and thus is comparable to the Seiberg-Witten map.

\par Employing this framework for Chern-Weil theory we introduced three types of Chern characters so that the third type, $\check{\mathrm{ch}}_*$, is shown to belong to $H_\star^*(X,\mathbb{Z})$ and has intimate correlation to Connes-Chern characters in cyclic (co)homology groups. On the other hand, $\check{\emph{\emph{ch}}}_*$ induces integral classes in $H_{\mathrm{dR}}^*(X,\mathbb{Z})$ which include significant information about topology of translation-invariant noncommutative Yang-Mills theories over $X$. Therefore, the topology of Abelian and topological anomalies of translation-invariant noncommutative Yang-Mills theories were studied thoroughly with correlation to Connes-Chern characters in cyclic (co)homology groups and the machinery of noncommutative geometry.

\section{Acknowledgments}
\par The author says his gratitude to S. Ziaee who was the main reason for appearing this article. Moreover, the author wishes to dedicate this work to Mohammad Reza Shajarian for all he has done to Iranian art and culture along last fifty years. Finally, my special thanks and highest regards go to the esteemed referee and the respectable editor of ROMP for all their honest considerations.










\end{document}